\documentclass[journal]{IEEEtran}
\usepackage[utf8]{inputenc}
\usepackage{url}
\usepackage{xcolor}
\usepackage{graphicx}
\usepackage{amsmath}
\DeclareMathOperator*{\argmax}{arg\,max}

\ifCLASSINFOpdf
\else
\fi
\usepackage{stackrel}
\begin{document}
%
\title{Laparoscopy Surgery CO$_2$ Removal via Generative Adversary Network and Dark Channel Prior}

\author{
Sebastián Salazar-Colores,
Hugo Alberto-Moreno,
Gerardo Flores$^*$,
and~César Javier Ortiz-Echeverri
\thanks{Sebastián Salazar-Colores,
Hugo Alberto-Moreno y
Gerardo Flores están con el Centro de Investigaciones en Óptica (CIO),
Lomas del Bosque 115, Lomas del Campestre, 37150 León, Gto (https://www.cio.mx).}

\thanks{César Javier Ortiz-Echeverri está con la Universidad Autónoma de Querétaro,
Facultad de informática, Av. de las Ciencias S/N, 76230 Juriquilla, Qro. (https://www.uaq.mx).}

\thanks{$^*$ Corresponding author. Email: gflores@cio.mx}
}

\maketitle

\begin{abstract}
Laparoscopic surgery uses a thin tube with a camera called a laparoscope, which is inserted into the abdomen through a small incision in the skin during surgery. This allows to a surgeon to see inside of the body without causing significant injury to the patient. These characteristics make laparoscopy a widely used technique.
In laparoscopic surgery, image quality can be severely degraded by surgical smoke caused by the use of tissue dissection tools which reduces the visibility of the observed organs and tissues. This lack of visibility increases the possibility of errors and surgery time with the consequences that this may have on the patient's health.
In this paper, we introduce a novel hybrid approach for computational smoke removal which is  based on the combination of a widely dehazing method used: the dark channel prior (DCP) and a pixel-to-pixel neural network approach: Generative Adversary Network (GAN). The experimental results have proven that the
proposed method achieves a better performance than the individual results of the DCP and GAN in terms of restoration quality, obtaining a PSNR value of 25 and SSIM index of 0.88 over a test set of synthetic images. 
\end{abstract}


%
\IEEEpeerreviewmaketitle

\section{Introducción}
\IEEEPARstart{L}{a} técnica de laparoscopia es ampliamente usada para la visualización y diagnóstico. Esta técnica consiste en la inserción de una cámara a través de pequeñas incisiones, en donde se insertan agujas para expandir el abdomen con gas de dióxido de carbono con el fin de dar cabida a otros instrumentos quirúrgicos. El uso de este gas inevitablemente disminuye la visibilidad de las cámaras usadas en esta técnica, esto puede causar errores en los algoritmos de procesamiento de imágenes así como reducir la visibilidad de los órganos y tejidos observados \cite{wang2018variational}.
Este inconveniente ha motivado la reciente investigación en procesamiento de imágenes, que se ha enfocado a la búsqueda de algoritmos para reducir los efectos del gas y por tanto, aumentar la visibilidad.

\begin{figure}[ht]
    \centering
    \includegraphics[width=\columnwidth]{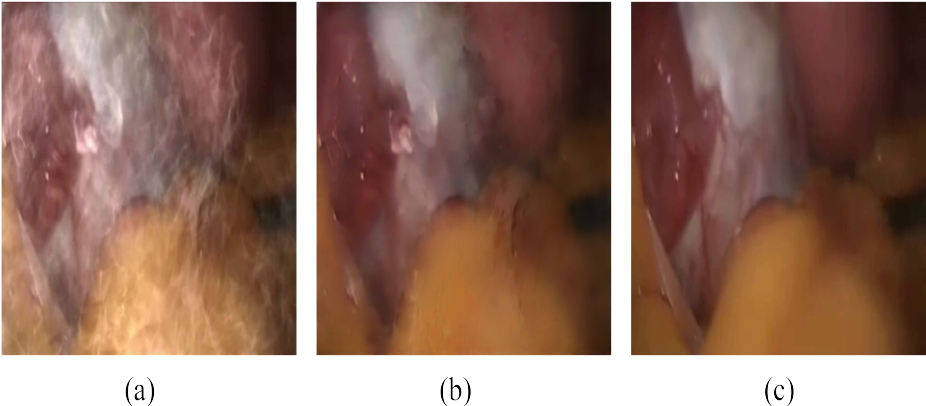}
    \caption{Funcionamiento del modelo DCP + GAN en una Imagen quirúrgica laparascopica. (a) Imágenes de entrada con CO$_2$ sintético, (b) nuestro resultado con DCP + GAN, (c) \textit{ground-truth}. }
    \label{fig:01}
\end{figure}

Una de las técnicas más usadas para este propósito parte del modelo de dispersión atmosférica, en donde se estima el mapa de transmisión y la luz atmosférica \cite{xu2015review}. Por ejemplo, en \cite{wang2018variational} se propone una técnica basada en el modelo físico de la dispersión del gas, similar al modelo atmosférico usado en el procesamiento de imágenes con niebla (\textit{dehazing}), en donde se estima el nivel de humo, considerando el hecho de que éste tiene un bajo contraste así como bajas diferencias entre canales. En base a esta observación, se define una función de costo y se resuelve utilizando un método Lagrangiano aumentado. En \cite{kotwal2016joint} se formula el problema de \textit{dehazing} y eliminación de humo en las imágenes laparoscópicas, como un problema de inferencia bayesiana, en donde es usado un sistema probabilístico con modelos de las imágenes sin niebla, así como el mapa de transmisión que indica la atenuación del color en presencia de humo. En \cite{gu2015virtual} se implementa un sistema digital basado en el canal oscuro, en donde se observó que estadísticamente las imágenes libres de niebla en exteriores contienen algunos pixeles con muy baja intensidad al menos en uno de los canales de color.

Muchos problemas en el procesamiento de imágenes, gráficos y visión implican la conversión de una imagen de entrada a una imagen de salida determinada. Para ello, a menudo se usan algoritmos específicos, pero que tienen en común el mapeo pixel a pixel (\textit{pix2pix}). Las redes generativas antagónicas (GAN) son una solución de propósito general que ha funcionado muy bien en una amplia variedad de aplicaciones que requieren un mapeo pixel a pixel, siendo actualmente el estado del arte en muchas de ellas \cite{isola2017image, karras2019style, chen2019qsmgan, zhang2019image, wang2018video, zhang2017adversarial}. Estas redes constan de dos partes principales, una función generadora, en donde a partir de una imagen de entrada, se genera una imagen de salida con las variaciones determinadas. Por otra parte, una función discriminante que evalúa la imagen generada, con una imagen real con el fin de clasificarla como real o falsa. La idea básica de las redes GAN es entonces lograr la generación de imágenes tan fieles a las originales que el discriminador no logre encontrar diferencia alguna. En el trabajo realizado por Isola et. al. \cite{isola2017image}, se demuestra que este enfoque es efectivo para sintetizar imágenes a partir de mapas de etiquetas, reconstruir objetos a partir de mapas de bordes y colorear imágenes originalmente en escala de grises. Este nuevo enfoque reduce considerablemente el ajuste de hiperparámetros, tal como pasa con las redes neuronales convolucionales (CNN), pues si bien el proceso de aprendizaje es automático, muchas veces la elección de la función de pérdida y la arquitectura dependen de un gran esfuerzo manual por parte del diseñador. En trabajos recientes se han propuesto arquitecturas basadas en aprendizaje profundo (\textit{Deep Learning}) para reducir el efecto del gas en imágenes laparoscópicas \cite{wang2019multiscale, bolkar2018deep}.
En \cite{chen2018unsupervised} se propone el uso de las redes GAN para el aprendizaje no supervisado y la remoción del gas en imágenes.

En este trabajo, como se muestra en la Figura \ref{fig:01}, dados los excelentes resultados obtenidos en la literatura con el canal oscuro \cite{wang2018smoke}, planteamos la posibilidad de utilizar el canal oscuro como una etapa de preprocesamiento para la red Neuronal antagónica Generativa (GAN), teniendo como hipótesis que dicha etapa previa reducirá la complejidad de los patrones presentes en las imágenes en comparación al enfoque \textit{end to end}.
Los experimentos fueron realizados sobre un conjunto de datos con imágenes de laparoscopia a las cuales se les ha generado artificialmente el efecto del gas $CO_2$ mediante \textit{software} de edición de vídeo. 
De acuerdo a las métricas usadas para evaluar los métodos propuestos, el uso de redes GAN para reducir el efecto del $CO_2$ mejora significativamente la reconstrucción de la imagen frente al enfoque basado en canal oscuro y modelo de dispersión. En cuanto a la combinación de dichos métodos, se encontró una ligera mejora en una de las métricas usadas.
El presente trabajo está estructurado de la siguiente forma: en la sección II, se presentan los fundamentos teóricos y conceptuales usados para la implementación de la red, así como las métricas para la evaluación de los resultados. En la sección III se explican las configuraciones usadas en el diseño del experimento.  En la sección IV se presentan los resultados obtenidos, así como las tablas comparativas respecto a trabajos relacionados. Finalmente, se presenta una discusión, las conclusiones y una perspectiva de futuras investigaciones.

\section{Marco teórico}

En esta sección se presentan los fundamentos teóricos necesarios para la implementación de la metodología propuesta. 

\subsection{Modelo de dispersión atmosférica}

La degradación atmosférica es un fenómeno físico causado por partículas en el medio atmosférico que absorben y dispersan la luz degradando así la adquisición de la imagen \cite{li2018benchmarking}. Esta degradación puede ser expresada utilizando el modelo dicromático \cite{gibson2013fast}, como
\begin{equation}\label{eq1}
    I(x,y)=J(x,y)t(x,y)+A(1-t(x,y)),
\end{equation}
en donde $I(x,y)$ es la intensidad de la imagen $RGB$ en cada pixel $(x,y)$ con presencia de niebla, $J(x,y)$ es la intensidad de la imagen $RGB$ en cada pixel $(x,y)$ sin degradación, $A$ es el vector normalizado de tres elementos que representan la luz atmosférica (color de la fuente de iluminación), por ejemplo, la luz blanca está representada por $A=[1 1 1]$ en el espacio de color $RGB$. Finalmente, $t(x,y)$ representa el mapa de transmisión en un atmósfera homogénea modelado por
\begin{equation}\label{eq2}
    t(x,y)=e^{-\beta d(x,y)}, 0<t(x,y)<1,
\end{equation}
donde $\beta$ es el coeficiente de dispersión, que depende principalmente del tamaño de la partícula de dispersión \cite{gibson2013fast}, y $d(x, y)$ representa la distancia de la cámara a la escena en la posición del píxel $(x,y)$. 

Utilizando el modelo dicromático de la Ecuación \ref{eq1}, es posible obtener la imagen sin los efectos de neblina $J(x,y)$ siempre y cuando sea posible obtener una estimación precisa de $t(x, y)$ y $A$ \cite{he2010single}.
%
\subsection{Canal oscuro}
El canal oscuro representado como $I^{dark}(x,y)$ se define como
\begin{equation}\label{eq3}
    I^{\textrm{dark}}(x,y) = \min_{c\in \{R,G,B\}}{\left({\min_{z\in \Omega\{x,y\}}} \frac{I^c(z)}{A^c}\right)} ,
\end{equation}
en donde $I^{dark}(x,y)$ es una matriz generalmente cuadrada, centrada en el pixel $(x,y)$; $\Omega(x,y)$ es un \textit{kernel} generalmente cuadrado centrado en la posición $(x,y)$; $I^c(z)$ son los elementos de la imagen $I$ en las posiciones $z\in\Omega(x,y)$; y $c$ representa cada imagen en los respectivos canales $R$, $G$ y $B$.
El principio del canal oscuro es definido como la relación estadística entre el canal oscuro y regiones sin neblina, donde se cumple que
\begin{equation}
    I^{\textrm{dark}}(x,y)\to 0.
\end{equation}
La relación existente entre $I^{\textrm{dark}}$ y $t$ se expresa como
\begin{equation}\label{eq4}
    t(x,y)=1-\omega I^{\textrm{dark}}(x,y),
\end{equation}
en donde $0<\omega<1$ representa el nivel de restauración. De acuerdo a \cite{he2010single}, $\omega=0.95$, en donde se argumenta que este valor produce un mejor aspecto en las imágenes restauradas. Por otra parte, la luz atmosférica puede ser estimada mediante la ecuación
\begin{equation}\label{eq5}
    A= \max\sum_{c=1}^{3}I^c \left( \stackbin[(x,y)\in 0.1\% (h,w)]{}{\argmax} \left(I^{dark}(x,y) \right)    \right),
\end{equation}
donde $h$ es la altura y $w$ es el ancho de la imagen.
%
\subsection{Redes Generativas Antagónicas}
Las Redes Generativas Antagónicas (GAN) presentadas en 2014 \cite{goodfellow2014generative}, son una clase de algoritmos de inteligencia artificial que se utilizan en el aprendizaje no supervisado, implementadas por un sistema de dos redes neuronales que compiten mutuamente. 
Las GAN se implementan usando capas convolucionales y deconvolucionales (convolucionales transpuestas) para el generador ($G$), el cual está diseñado para producir salidas que no puedan ser distinguidas de las imágenes reales por una persona entrenada. Por otro lado, una red neuronal convolucional (CNN) constituye el discriminador ($D$), capacitado para hacer lo mejor posible en la detección de las falsificaciones realizadas por G. Los elementos fundamentales para la implementación de una red GAN se mencionan a continuación.

\subsubsection{Capas convolucionales}

Una capa convolucional está compuesta por varios \textit{kernel} de convolución que se utilizan para calcular diferentes mapas de características. Específicamente, cada neurona de un mapa de características está conectada a una región de neuronas vecinas en la capa anterior. El nuevo mapa de características se puede obtener haciendo una convolución de la entrada con un kernel aprendido y luego aplicando una función de activación no lineal.
Para generar cada mapa de características, el \textit{kernel} es compartido por todas las ubicaciones espaciales de la entrada. Los mapas de características completos se obtienen utilizando \textit{kernel} diferentes. Matemáticamente, el valor de la característica en la ubicación $(i,j)$ en el mapa de característica $k$-th de la capa $l$-th, $z^l_{i,j,k}$, se calcula como
\begin{equation}\label{eq6}
    z^l_{i,j,k} =  {\mathrm{w}^l_k }^T \mathrm{x}^l_{i,j} + b^l_k
\end{equation}
donde $\mathrm{w}^l_k$ y $b^l_k$ son el vector de peso y el término de sesgo del filtro $k$-th de la capa $l$-th respectivamente; y $\mathrm{x}^l_{i,j}$ es el parche de entrada centrado en la ubicación $(i, j)$ de la capa $l$-th. El \textit{kernel} $\mathrm{w}^l_k$ que genera el mapa de características $z^l _{:,:,k}$ es compartido. Tal mecanismo de reparto de peso tiene varias ventajas, tales como que puede reducir la complejidad del modelo y hacer que la red sea más fácil de entrenar.
\subsubsection{ReLU}
La unidad lineal rectificada (\textit{ReLU}) es una de las funciones de activación no saturada más notable. La función de activación \textit{ReLU} se define como:
\begin{equation}\label{eq7}
    a_{i,j,k}= \max(z_{i,j,k},0)\
\end{equation}
donde $z_{i,j,k}$ es la entrada de la función de activación en la ubicación $(i,j)$ en el canal $k$-th.  \textit{ReLU} es una función lineal por partes que poda la parte negativa a cero y retiene la parte positiva. La simple operación $\max(\cdot)$ de \textit{ReLU} le permite computar mucho más rápido que las funciones de activación sigmoide o $\tanh{(\cdot)}$, y también induce la dispersión en las unidades ocultas y permite a la red obtener fácilmente representaciones dispersas. Se ha demostrado que las redes profundas pueden ser entrenadas eficientemente usando \textit{ReLU} incluso sin necesidad de pre-entrenamiento \cite{krizhevsky2012imagenet}

\subsection{Redes Generativas Antagónicas Condicionales}

Las GAN son modelos generativos que aprenden un mapeo a partir de un vector de ruido aleatorio $z$ para una imagen de salida $y$, $G:z \rightarrow y$. Por su parte, las redes condicionales GAN, aprenden a partir de la imagen de entrada $x$ y el vector de ruido aleatorio $z$, $G:\{ x,z \} \rightarrow y$ \cite{isola2017image}.
Las capas deconvolucionales o convolucionales transpuestas de una CNN, han hecho posible la generación de una salida del mismo tamaño que una imagen de entrada. Sin embargo, la pérdida $L_1$ o $L_2$ utilizada como métrica de similitud lleva a una predicción de imágenes borrosas. las redes condicionales GAN buscan resolver este problema añadiendo una pérdida adversa implementada como una \textit{Convolutional Neural Netwok} (CNN) separada, que posee un sistema binario de salida que funciona como clasificador.\\

\section{Materiales y métodos}

\subsection{Obtención de la base de datos}

Los videos de operaciones quirúrgicas de los que se extrajeron las imágenes utilizadas en el presente trabajo fueron obtenidos de repositorios públicos \cite{MEDtube} así como vídeos proporcionados por un grupo de médicos especialistas en cirugías laparoscópicas \footnote{Un video comparativo de los resultados de los métodos propuestos de un video  quirúrgico real puede ser consultado en \url{https://youtu.be/QvUKcHonCHw}}. De los vídeos disponibles se obtuvieron $6000$ imágenes representativas de diferentes niveles de afectaciones por $CO_2$. Para generar los datos de entrada, se simuló artificialmente el gas, usando el \textit{software} de procesamiento gráfico de código abierto \textit{Blender}, formando imágenes de entrada con dimensiones de $512\times512$ pixeles.

\subsection{Métricas}

Para tener una visión del desempeño del método propuesto en este trabajo se evalúan los resultados empleando métricas ampliamente utilizadas en la literatura: la Proporción máxima de señal a ruido PSNR y el índice de similitud estructural SSIM, enseguida se explicarán brevemente estos conceptos:\\ 

\begin{itemize}
\item 
La Relación Señal a Ruido de Pico o PSNR (del inglés \textit{Peak Signal-to-Noise Ratio}) es una medida cuantitativa de la calidad de una reconstrucción \cite{huynh2008scope}. Es utilizada ampliamente en imágenes. Para definir la métrica PSNR es necesario definir el error cuadrático medio (del inglés \textit{MSE}), el cuál para dos imágenes monocromas $I$ y $J$ de tamaño $m\times n$ se define como
\begin{equation}
\textrm{MSE}=\frac{1}{mn}\sum_{i=0}^{m-1}\sum_{j=0}^{n-1}||I(i,j)-J(i,j)||^2
\end{equation}
y el $\textrm{PSNR}$ está dado por
\begin{equation}
\textrm{PSNR}=10 \log_{10}\Bigl(\frac{\textrm{MAX}^2_I}{\textrm{MSE}}\Bigr)=20 \log_{10}\Bigl(\frac{\textrm{MAX}_I}{\sqrt{\textrm{MSE}}}\Bigr)
\end{equation}
donde $\textrm{MAX}=2^B-1$ y $B$ es el número de Bits utilizados en la imagen. Valores altos de $\textrm{PSNR}$ indican mejores restauraciones.\\
    
\item Índice SSIM es una métrica de similitud de imagen perceptiva que fue propuesta como alternativa al error medio cuadrático (MSE) y PSNR para aumentar la correlación con la evaluación subjetiva.
Para las imágenes originales y reconstruidas $I$ y $J$, SSIM se define como
\begin{equation}
    \textrm{SSIM}(I,J)= \frac{(2\mu_I\mu_J+C_1 )(2\sigma_{IJ}+C_2 )}{(\mu_I^2+\mu_J^2+C_1 )(\sigma_I^2+\sigma_J^2+C_2 ) }
\end{equation}
con $\mu$, $\sigma$ y $\sigma_{IJ}$, como la media la varianza y la covarianza de las imágenes, respectivamente.
\end{itemize}
%
\subsection{Método propuesto}
\begin{figure}[h]
\centering
\includegraphics[width=\columnwidth]{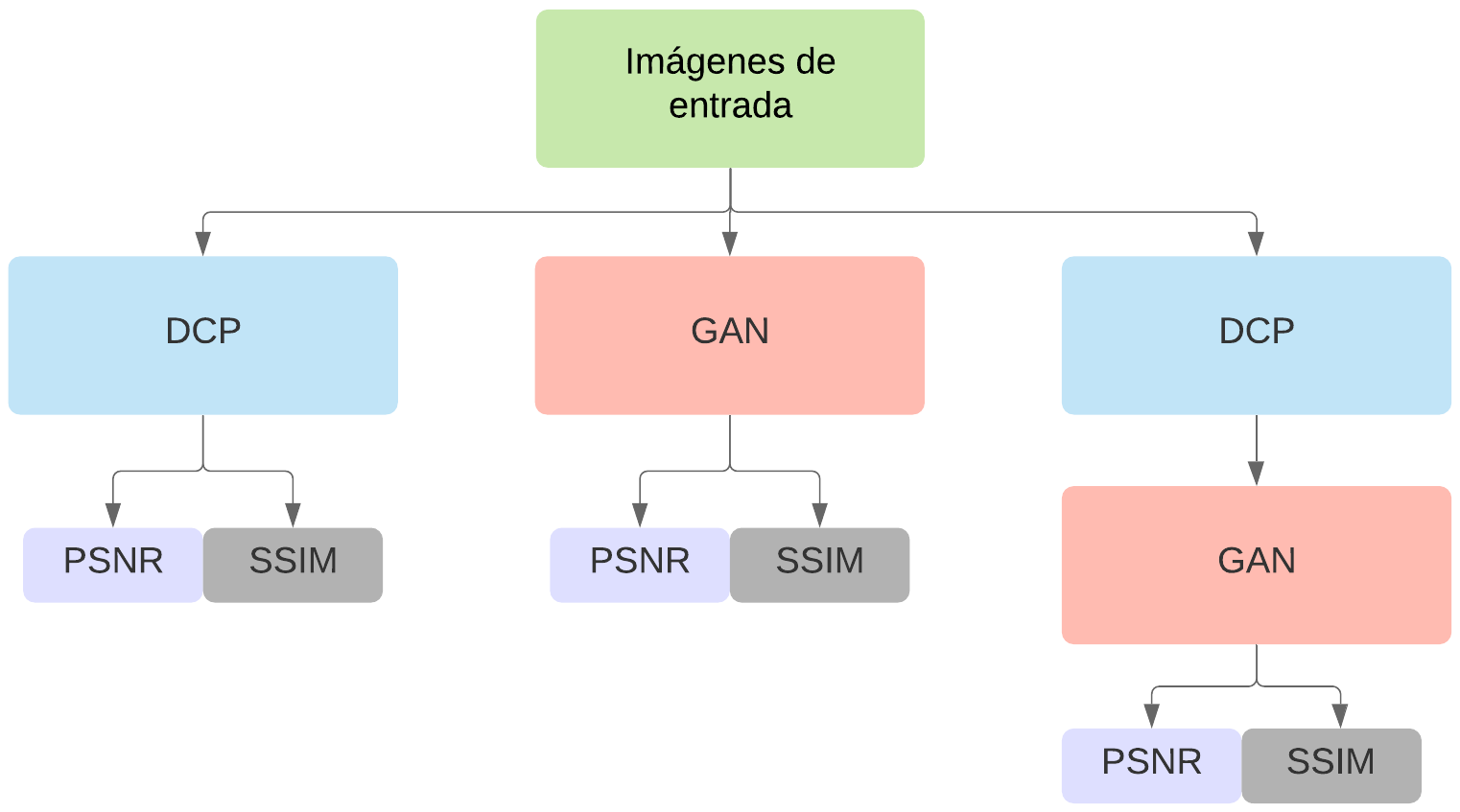}
\caption{Metodología propuesta para la evaluación de desempeño.}
\label{METHOD}
\end{figure}

El método propuesto está basado en el supuesto que una red neuronal GAN tiene un mejor rendimiento en tanto su entrada sea más parecida a su salida esperada. Entonces al aplicar el canal oscuro previamente a entrenar la red neuronal, el rendimiento de la red neuronal se verá incrementado. Para realizar un análisis del rendimiento del enfoque propuesto, tres experimentos fueron propuestos:
\begin{itemize}
    \item Técnica basada en el canal oscuro. Como se ha comentado previamente el canal oscuro ha dado excelentes resultados en tareas de dehazing, no obstante el problema abordado es distinto  ya que la atmósfera no es homogénea por lo que ya no existe una relación entre el canal oscuro y la distancia de los objetos. Sin embargo la distancia ($d(x)$) de la cámara al cuerpo humano es depreciable, por tanto aqui el canal oscuro tiene una correspondencia directa con el coeficiente de dispersión $\beta$. Es entonces que el canal oscuro tiene tambien validez para este problema. Describiendo lo anteriormente mencionado de forma explícita, tenemos que:\\
    
    De \eqref{eq4} tenemos que la relación entre el canal oscuro y la transmisión es como sigue
    \begin{equation}\label{eq11}
    t(x,y)=1-\omega I^{\textrm{dark}}(x,y).
    \end{equation}
    Luego, sustituyendo \eqref{eq1} en \eqref{eq4} y considerando además que la atmósfera no es homogénea, se obtienen diferentes coeficientes de dispersión para cada elemento $(x,y)$ como sigue
    \begin{equation}\label{eq12}
    e^{-\beta(x,y) d(x,y)}=1-\omega I^{\textrm{dark}}(x,y).
    \end{equation}
    Por tanto se sigue que
    \begin{equation}\label{eq13}
    I^{\textrm{dark}}(x,y)=\frac{1-e^{-\beta(x,y) d(x,y)}}{\omega}.
    \end{equation}
    Luego, suponiendo la distancia del cuerpo  constante y $\omega=1$, esto es
    \begin{equation}\label{eq14}
    I^{\textrm{dark}}(x,y)=1-e^{-d\beta(x,y)},
    \end{equation}
    podemos concluir que la relación del canal oscuro con la transmisión esta en función ahora del coeficiente y densidad de partículas $\beta$ en cada uno de los píxeles, resultando útil para la eliminación de los efectos.

    En la Figura \ref{DCP} se muestra detalles del método basado en el DCP implementado.
    
    \begin{figure}[h]
    \centering
    \includegraphics[width=\columnwidth]{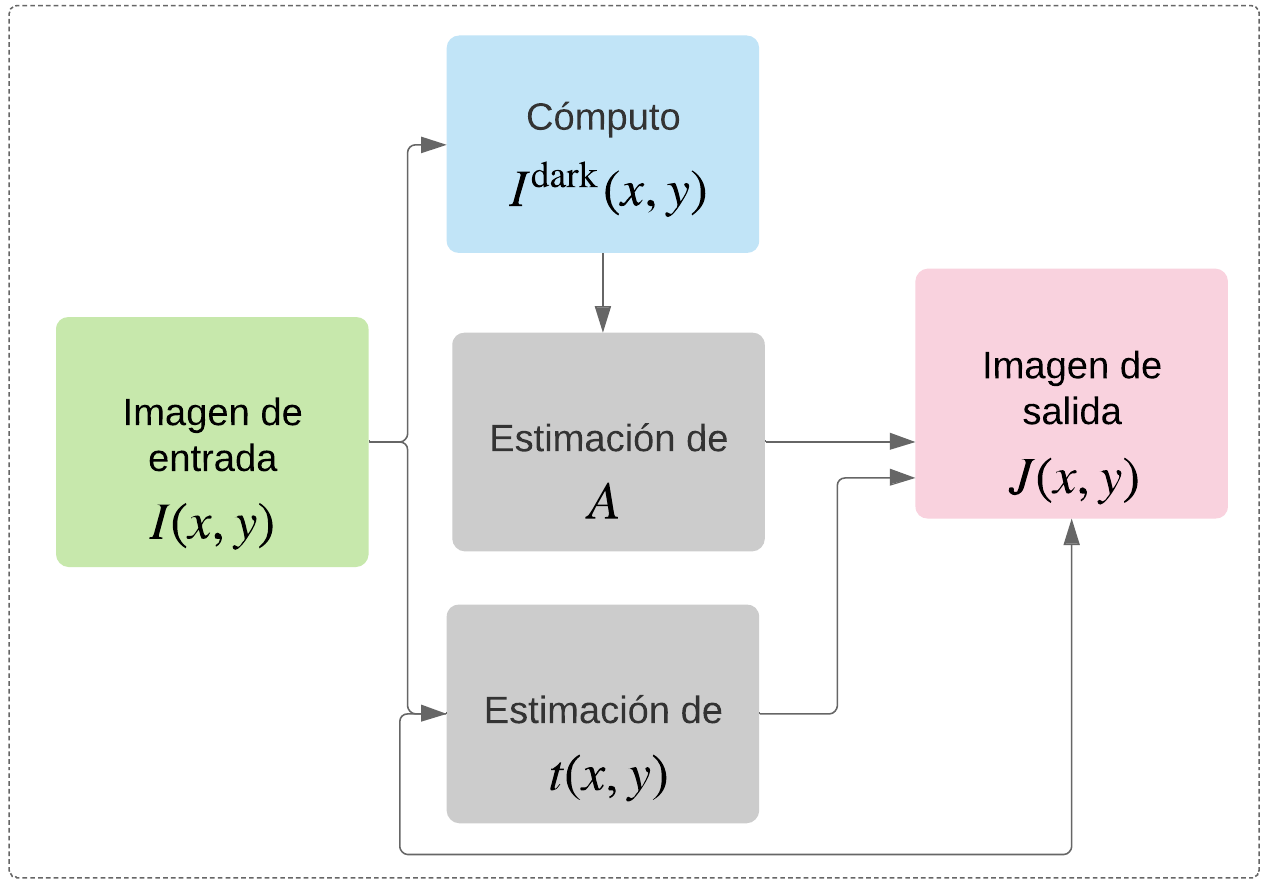}
    \caption{Método basado en DCP.}
    \label{DCP}
    \end{figure}

    \item Utilizando la red Neuronal Generativa Antagónica propuesta. Se muestra un diagrama de la red neuronal utilizada en la Figura \ref{GAN}. En las Tablas \ref{tab:generadora} y \ref{tab:discriminadora} se  muestra  la  arquitectura  y  los  hiper-parámetros del generador y discriminador empleado. Como función de optimización en la red neuronal se empleo el Momentum Adaptable (ADAM). Este algoritmo es una extensión del descenso de gradiente estocástico para actualizar los pesos de red de forma iterativa en función de los datos de entrenamiento, recientemente ha visto una adopción más amplia para aplicaciones de aprendizaje profundo en visión por computadora y procesamiento de lenguaje natural. Como métrica de pérdida se utiliza MSE que es el promedio del error al cuadrado este se usa como la función de pérdida para la regresión.
    \begin{figure}[ht]
\centering
\includegraphics[width=\columnwidth]{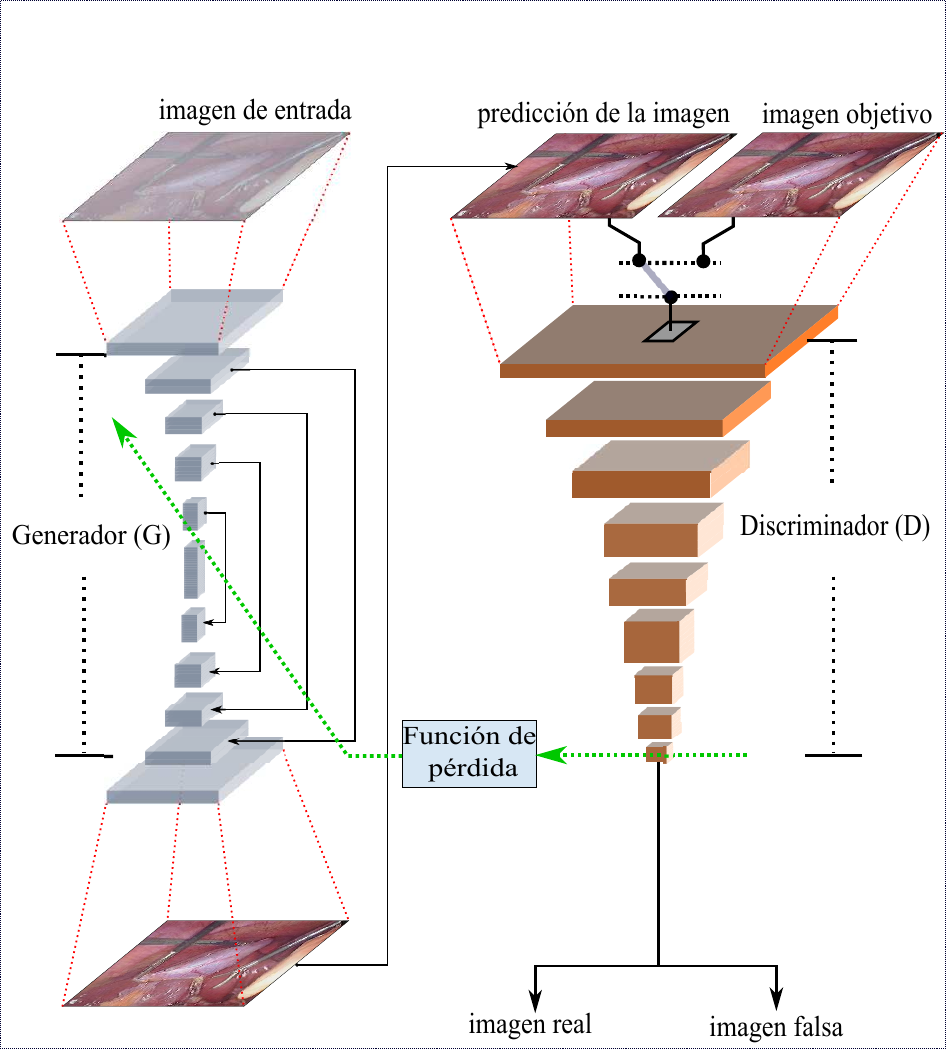}
\caption{Diagrama arquitectura GAN. La imagen de entrada pasa a través de varias capas convolucionales hasta lograr una reducción de dimensionalidad en la capa intermedia de la estructura generadora (G), seguidamente se realiza un proceso inverso con capas deconvolucionales, hasta llegar a una imagen con igual dimensionalidad a la imagen de entrada la cual será la predicción de la imagen sin ruido ($CO_2$). Por otra parte, un discriminador ($D$) basado en operaciones convolucionales compara la predicción de la imagen, con la imagen objetivo.}
\label{GAN}
\end{figure}
    \item Usando una combinación de los enfoques antes mencionados, en donde las imágenes de entrada en la Figura \ref{GAN}, son las imágenes de salida $J$ de la Figura \ref{DCP}. La evaluación del desempeño para cada caso es representado en la Figura \ref{METHOD}.
\end{itemize}

%
\begin{center}
\begin{table}[!t]
\renewcommand{\arraystretch}{1.3}
\caption{Arquitectura de la red neuronal Generadora utilizada.}
\label{tab:generadora}
\centering
\tiny	\begin{tabular}{ccclcc}
\hline
\textbf{Capa} & \textbf{Conv} & \textbf{Kernel} & \textbf{Salto} & \textbf{Definición}                                                                    & \textbf{Tamaño}           \\ \hline\hline
1    & 64          & 4      & 2     & Conv -\textgreater Batchnorm -\textgreater Leaky ReLU                   & (256, 256, 64)   \\ \hline
2    & 128         & 4      & 2     & Conv -\textgreater Batchnorm -\textgreater Leaky ReLU                   & (128, 128, 128)  \\ \hline
3    & 256         & 4      & 2     & Conv -\textgreater Batchnorm -\textgreater Leaky ReLU                   & (64, 64, 256)    \\ \hline
4    & 512         & 4      & 2     & Conv -\textgreater Batchnorm -\textgreater Leaky ReLU                   & (32, 32, 512)    \\ \hline
5    & 512         & 4      & 2     & Conv -\textgreater Batchnorm -\textgreater Leaky ReLU                   & (16, 16, 512)    \\ \hline
6    & 512         & 4      & 2     & Conv -\textgreater Batchnorm -\textgreater Leaky ReLU                   & (8, 8, 512)      \\ \hline
7    & 512         & 4      & 2     & Conv -\textgreater Batchnorm -\textgreater Leaky ReLU                   & (4, 4, 512)      \\ \hline
8    & 512         & 4      & 2     & Conv -\textgreater Batchnorm -\textgreater Leaky ReLU                   & (1, 1, 512)      \\ \hline
9    & 512         & 4      & 2     & Deconv -\textgreater Batchnorm-\textgreater{}ReLU-\textgreater{}Dp(0.5) & (2, 2, 1024)     \\ \hline
10   & 512         & 4      & 2     & Deconv -\textgreater Batchnorm-\textgreater{}ReLU-\textgreater{}Dp(0.5) & ( 4,4, 1024)     \\ \hline
11   & 512         & 4      & 2     & Deconv -\textgreater Batchnorm-\textgreater{}ReLU-\textgreater{}Dp(0.5) & ( 8,8, 1024)     \\ \hline
12   & 512         & 4      & 2     & Deconv -\textgreater Batchnorm-\textgreater{}ReLU-\textgreater{}Dp(0.5) & ( 16, 16, 1024)  \\ \hline
13   & 512         & 4      & 2     & Deconv -\textgreater Batchnorm-\textgreater{}ReLU                       & ( 32, 32, 1024)  \\ \hline
14   & 256         & 4      & 2     & Deconv -\textgreater Batchnorm-\textgreater{}ReLU                       & ( 64, 64, 512)   \\ \hline
15   & 128         & 4      & 2     & Deconv -\textgreater Batchnorm-\textgreater{}ReLU                       & ( 128, 128, 256) \\ \hline
16   & 64          & 4      & 2     & Deconv -\textgreater Batchnorm-\textgreater{}ReLU                       & ( 256, 256, 128) \\ \hline
17   & 3           & 4      & 2     & tanh                                                                    & (512, 512, 3)    \\ \hline
\end{tabular}
\end{table}
\label{tabla1}
\end{center}

\begin{center}
    
\begin{table}[!t]
\renewcommand{\arraystretch}{1.3}
\caption{Arquitectura de la red neuronal Discriminadora utilizada.}
\label{tab:discriminadora}
\tiny
\center
\begin{tabular}{cccclc}
\hline
\textbf{Capa} & \textbf{Conv} & \textbf{kernel} & \textbf{Saltos} & \textbf{Definición}                                              & \textbf{Tamaño}         \\ \hline\hline
1             & 64          & 4      & 2      & (Conv -\textgreater BatchNorm -\textgreater Leaky ReLU) & (128, 128, 64) \\ \hline
2             & 128         & 4      & 2      & (Conv -\textgreater BatchNorm -\textgreater Leaky ReLU) & (64, 64, 128)  \\ \hline
3             & 256         & 4      & 2      & (Conv -\textgreater BatchNorm -\textgreater Leaky ReLU) & (32, 32, 256)  \\ \hline
4             & 0           & 0      & 0      & (ZeroPadding2D)                                         & (34, 34, 256)  \\ \hline
5             & 512         & 4      & 1      & (Conv )                                                 & (31, 31, 512)  \\ \hline
6             & 0           & 0      & 0      & (BatchNorm -\textgreater Leaky ReLU-\textgreater ZeroPadding)        & (33, 33, 512)  \\ \hline
7             & 1           & 4      & 1      & (Conv)                                                  & (30, 30, 1)    \\ \hline
\end{tabular}
\end{table}
\label{tabla2}
\end{center}

En la Figura \ref{METHOD} se muestra la metodología propuesta para evaluar el desempeño de las tres configuraciones mencionadas.
%

\subsection{Configuración de los experimentos}
Los experimentos fueron realizados sobre una computadora con un procesador Ryzen Threaddriper 1900, 128 Gb de memoria Ram y una tarjeta Gráfica Nvidia RTX 2080 Ti, usando Linux Ubuntu 18.10, Python 3.5, Librerias OpenCV 3.3 y Keras 2.1.0.

\newpage
\section{Resultados}

\subsection{Comparación cuantitativa}

\begin{figure}[h]
    \centering
    \includegraphics[width=0.45\textwidth]{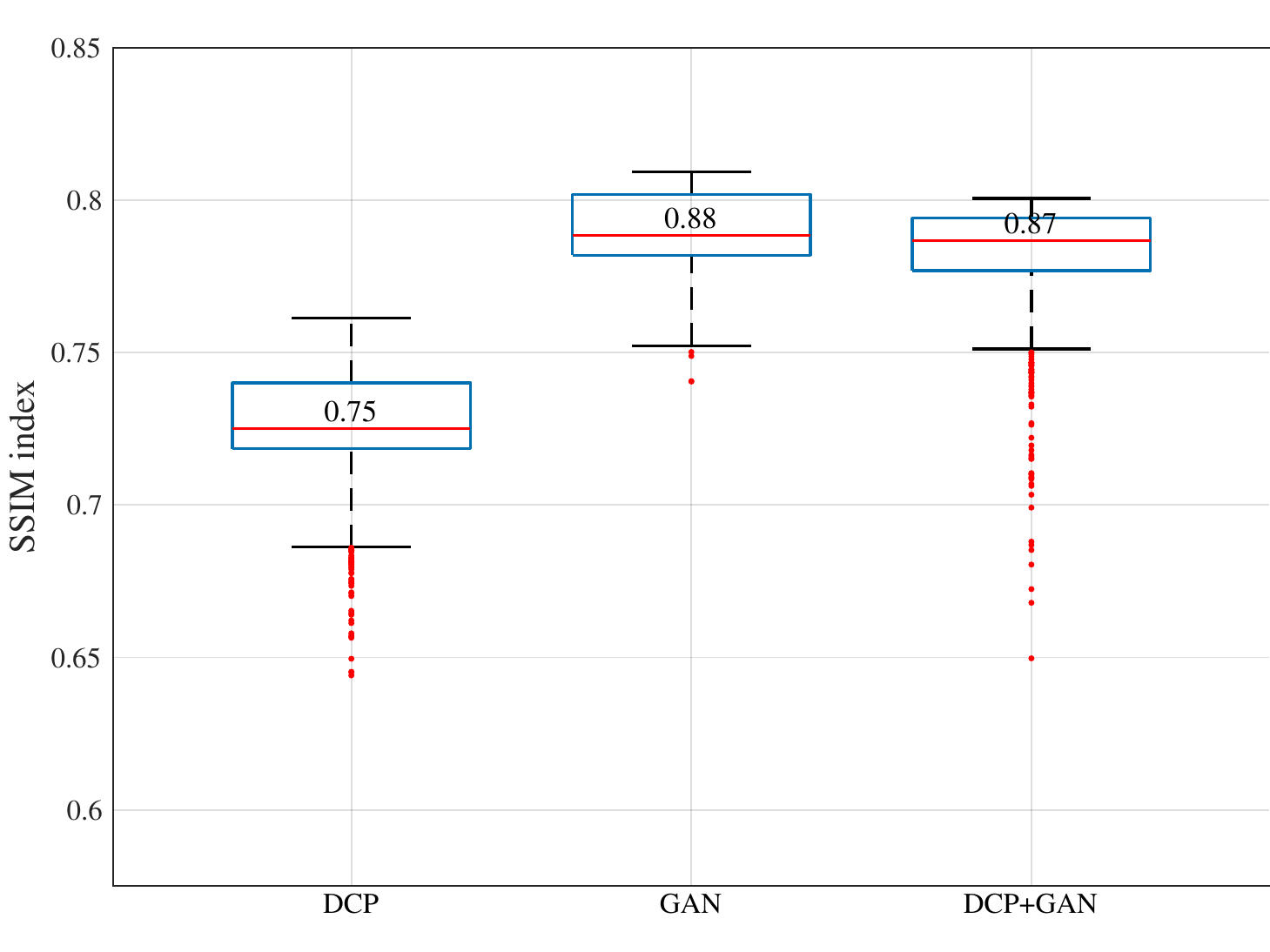}
    \caption{Comparativa del desempeño de los métodos propuestos de acuerdo al índice SSIM.}
    \label{boxplot1}
\end{figure}

\begin{figure}[h]
    \centering
    \includegraphics[width=0.45\textwidth]{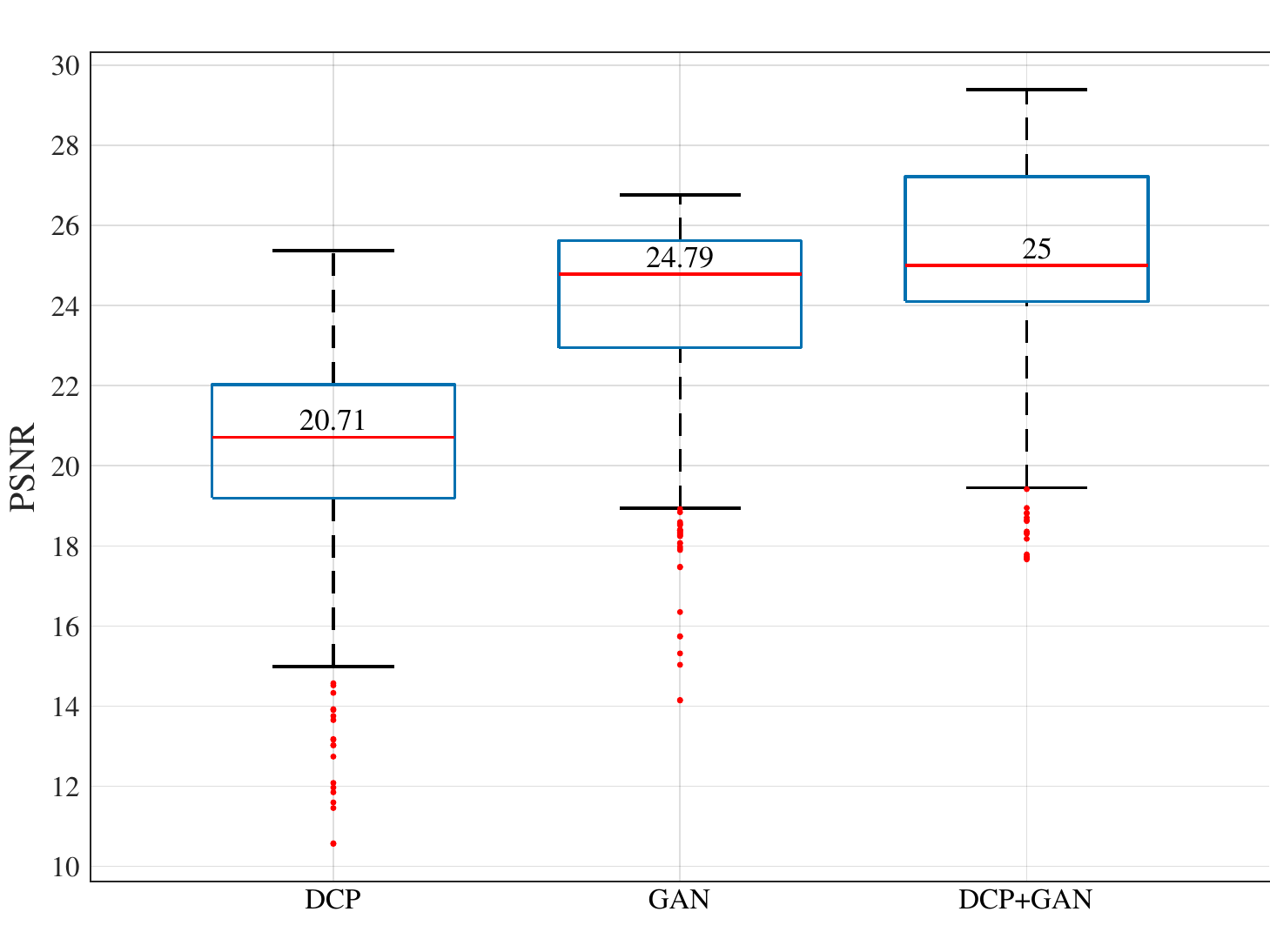}
    \caption{Comparativa del desempeño de los métodos propuestos de acuerdo al PSNR.}
    \label{boxplot2}
\end{figure}

En la Figura \ref{boxplot1} y \ref{boxplot2} se muestra la comparación de las métricas SSIM y PSNR para los métodos DCP, GAN y DCP-GAN. La primera observación importante en estas figuras es la mejora significativa de los métodos GAN y DCP-GAN, frente a la reconstrucción basada únicamente en DCP, el cual alcanzó valores de $\textrm{SSIM}=0.75$ y $\textrm{PSNR}=20.71$, mientras GAN alcanzó valores de $\textrm{SSIM}=0.88$ y $\textrm{PSNR}=24.79$. En cuanto al desempeño entre GAN y DCP-GAN, se obseva que la métrica de PSNR fue ligeramente superior para DCP-GAN, con un valor $\textrm{PSNR}=25$, frente a $\textrm{PSNR}=24.79$ logrado por GAN. En el caso de la métrica SSIM, GAN obtuvo un valor medio de $\textrm{SSIM}=0.88$ frente a $\textrm{SSIM}=0.87$ de DCP-GAN. Teniendo en cuenta la definición de SSIM, diseñada para medir parámetros perceptuales propios del ojo humano, este resultado podría deberse a una saturación generada por la etapa DCP.

\clearpage

\begin{figure*}
    \centering
    \includegraphics[width=\textwidth]{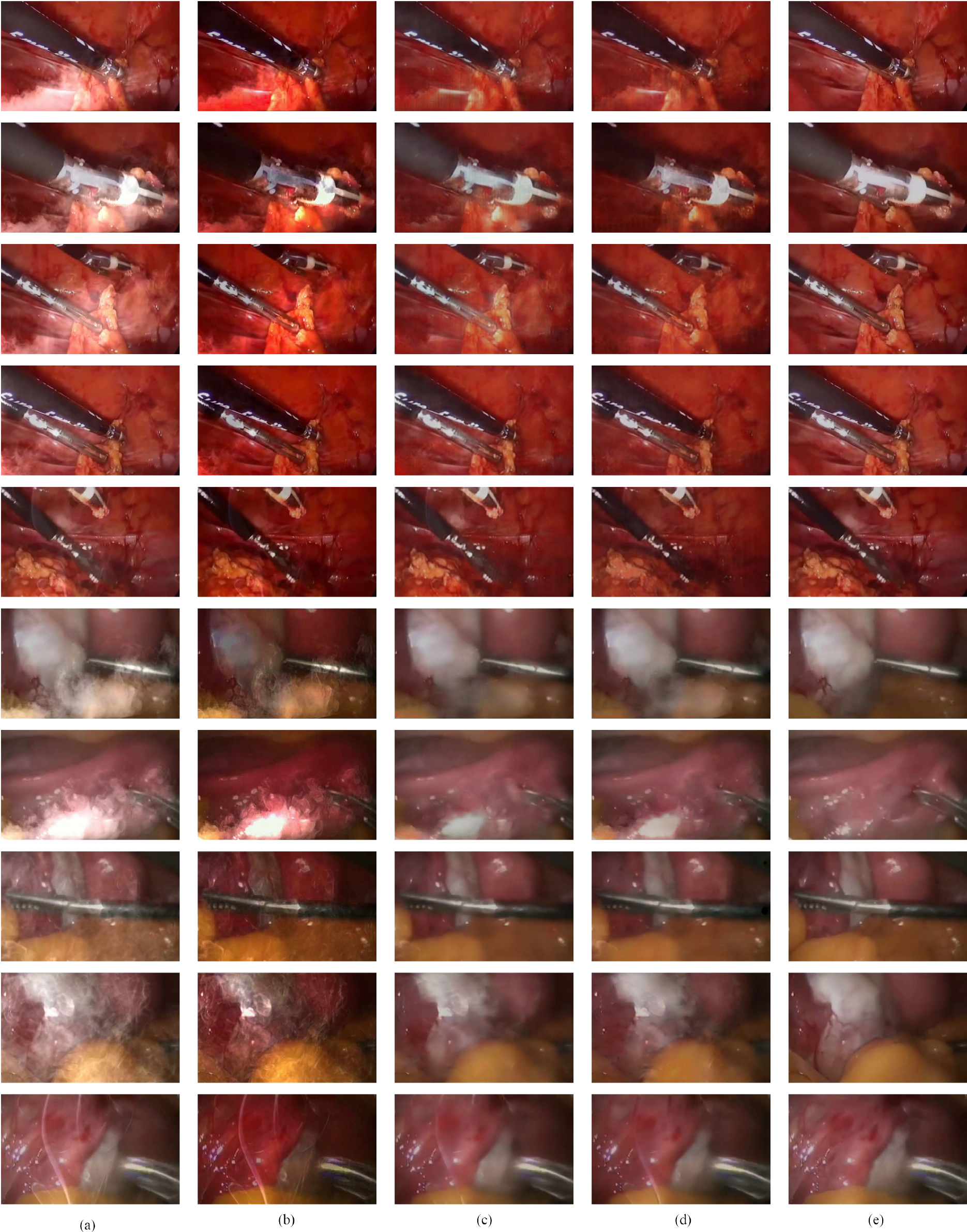}
    \caption{Comparación de resultados con los métodos utilizados. (a) Imágenes de entrada con CO$_2$ sintético,  (b) Dark Chanel Prior, (c) GAN, (d) DCP + GAN, (e) \textit{Ground-truth}.}
    \label{}
\end{figure*}

\newpage
\clearpage

\section{Conclusiones}
En este articulo fue propuesto un método para mejorar la reducida visibilidad en imágenes laparascopicas causada por la obstrucción visual debido al gas $CO_2$ suministrado durante cirugías. El método propuesto está basado en la combinación de dos etapas, la primera de ellas está centrada en el concepto de canal oscuro, con el cual se logra aumentar la visibilidad, posteriormente y en la segunda etapata se agrega un procesamiento utilizando una red generativa antagónica GAN. El método propuesto es capaz de reducir influencia del $CO_2$ en la imagen de salida y recuperar la gama de colores original, obteniendo una visión mas aproximada a la imagen real. Esta combinación de técnicas ha mostrado tener un alto desempeño en la reconstrucción de las imágenes objetivo, lo anterior de acuerdo a las métricas PSNR e índice SSIM usadas en el presente trabajo. 

Como trabajo futuro, se espera mejorar el comportamiento de la GAN al agregar una función de pérdida especialmente diseñada a la resolución del problema expuesto. Por otra parte se plantea hacer una búsqueda exhaustiva de los parámetros en el modelo de dispersión con el fin de ajustar la etapa de preprocesamiento en la entrada de la GAN. Así mismo, se espera optimizar la red para aplicaciones en tiempo real. Se continuará trabajando con médicos cirujanos a fin de proponer una solución en tiempo real que elimine el efecto visual del CO$_2$ en el vídeo captado por los médicos cirujanos.


%



\section*{Agradecimientos}

Hugo Alberto Moreno desea agradecer especialmente al CONACYT (Consejo Nacional de Ciencia y Tecnología) por el apoyo financiero brindado para sus estudios de maestría.

Los autores agradecen al médico Luis Alberto Tavares-de la Paz, Cirujano Oncólogo del Hospital Regional de Alta Especialidad del Bajío en León Guanajuato, por proporcionar el vídeo utilizado en los experimentos presentados en este trabajo.

\ifCLASSOPTIONcaptionsoff
  \newpage
\fi


\bibliography{references}

\begin{thebibliography}{10}

\bibitem{wang2018variational}
C.~Wang, F.~A. Cheikh, M.~Kaaniche, A.~Beghdadi, and O.~J. Elle, ``Variational
  based smoke removal in laparoscopic images,'' {\em Biomedical engineering
  online}, vol.~17, no.~1, p.~139, 2018.

\bibitem{xu2015review}
Y.~Xu, J.~Wen, L.~Fei, and Z.~Zhang, ``Review of video and image defogging
  algorithms and related studies on image restoration and enhancement,'' {\em
  Ieee Access}, vol.~4, pp.~165--188, 2015.

\bibitem{kotwal2016joint}
A.~Kotwal, R.~Bhalodia, and S.~P. Awate, ``Joint desmoking and denoising of
  laparoscopy images,'' in {\em 2016 IEEE 13th International Symposium on
  Biomedical Imaging (ISBI)}, pp.~1050--1054, IEEE, 2016.

\bibitem{gu2015virtual}
L.~Gu, P.~Liu, C.~Jiang, M.~Luo, and Q.~Xu, ``Virtual digital defogging
  technology improves laparoscopic imaging quality,'' {\em Surgical
  innovation}, vol.~22, no.~2, pp.~171--176, 2015.

\bibitem{isola2017image}
P.~Isola, J.-Y. Zhu, T.~Zhou, and A.~A. Efros, ``Image-to-image translation
  with conditional adversarial networks,'' in {\em Proceedings of the IEEE
  conference on computer vision and pattern recognition}, pp.~1125--1134, 2017.

\bibitem{karras2019style}
T.~Karras, S.~Laine, and T.~Aila, ``A style-based generator architecture for
  generative adversarial networks,'' in {\em Proceedings of the IEEE Conference
  on Computer Vision and Pattern Recognition}, pp.~4401--4410, 2019.

\bibitem{chen2019qsmgan}
Y.~Chen, A.~Jakary, C.~P. Hess, and J.~M. Lupo, ``Qsmgan: Improved quantitative
  susceptibility mapping using 3d generative adversarial networks with
  increased receptive field,'' {\em arXiv preprint arXiv:1905.03356}, 2019.

\bibitem{zhang2019image}
H.~Zhang, V.~Sindagi, and V.~M. Patel, ``Image de-raining using a conditional
  generative adversarial network,'' {\em IEEE Transactions on Circuits and
  Systems for Video Technology}, 2019.

\bibitem{wang2018video}
T.-C. Wang, M.-Y. Liu, J.-Y. Zhu, G.~Liu, A.~Tao, J.~Kautz, and B.~Catanzaro,
  ``Video-to-video synthesis,'' {\em arXiv preprint arXiv:1808.06601}, 2018.

\bibitem{zhang2017adversarial}
Y.~Zhang, Z.~Gan, K.~Fan, Z.~Chen, R.~Henao, D.~Shen, and L.~Carin,
  ``Adversarial feature matching for text generation,'' in {\em Proceedings of
  the 34th International Conference on Machine Learning-Volume 70},
  pp.~4006--4015, JMLR. org, 2017.

\bibitem{wang2019multiscale}
C.~Wang, A.~K. Mohammed, F.~A. Cheikh, A.~Beghdadi, and O.~J. Elle,
  ``Multiscale deep desmoking for laparoscopic surgery,'' in {\em Medical
  Imaging 2019: Image Processing}, vol.~10949, p.~109491Y, International
  Society for Optics and Photonics, 2019.

\bibitem{bolkar2018deep}
S.~Bolkar, C.~Wang, F.~A. Cheikh, and S.~Yildirim, ``Deep smoke removal from
  minimally invasive surgery videos,'' in {\em 2018 25th IEEE International
  Conference on Image Processing (ICIP)}, pp.~3403--3407, IEEE, 2018.

\bibitem{chen2018unsupervised}
L.~Chen, W.~Tang, and W.~John, ``Unsupervised learning of surgical smoke
  removal from simulation,'' in {\em 11th Hamlyn Symposium on Medical
  Robotics}, Bournemouth University, Fern Barrow, Poole, Dorset, BH12 5BB, UK,
  2018.

\bibitem{wang2018smoke}
C.~Wang, F.~A. Cheikh, M.~Kaaniche, and O.~J. Elle, ``A smoke removal method
  for laparoscopic images,'' {\em arXiv preprint arXiv:1803.08410}, 2018.

\bibitem{li2018benchmarking}
B.~Li, W.~Ren, D.~Fu, D.~Tao, D.~Feng, W.~Zeng, and Z.~Wang, ``Benchmarking
  single-image dehazing and beyond,'' {\em IEEE Transactions on Image
  Processing}, vol.~28, no.~1, pp.~492--505, 2018.

\bibitem{gibson2013fast}
K.~B. Gibson and T.~Q. Nguyen, ``Fast single image fog removal using the
  adaptive wiener filter,'' in {\em 2013 IEEE International Conference on Image
  Processing}, pp.~714--718, IEEE, 2013.

\bibitem{he2010single}
K.~He, J.~Sun, and X.~Tang, ``Single image haze removal using dark channel
  prior,'' {\em IEEE transactions on pattern analysis and machine
  intelligence}, vol.~33, no.~12, pp.~2341--2353, 2010.

\bibitem{goodfellow2014generative}
I.~Goodfellow, J.~Pouget-Abadie, M.~Mirza, B.~Xu, D.~Warde-Farley, S.~Ozair,
  A.~Courville, and Y.~Bengio, ``Generative adversarial nets,'' in {\em
  Advances in neural information processing systems}, pp.~2672--2680, 2014.

\bibitem{krizhevsky2012imagenet}
A.~Krizhevsky, I.~Sutskever, and G.~E. Hinton, ``Imagenet classification with
  deep convolutional neural networks,'' in {\em Advances in neural information
  processing systems}, pp.~1097--1105, 2012.

\bibitem{MEDtube}
``Medtube.'' \url{https://medtube.es/}.
\newblock Accessed: 2019-09-09.

\bibitem{huynh2008scope}
Q.~Huynh-Thu and M.~Ghanbari, ``Scope of validity of psnr in image/video
  quality assessment,'' {\em Electronics letters}, vol.~44, no.~13,
  pp.~800--801, 2008.

\end{thebibliography}
\bibliographystyle{ieeetr} 

%

\begin{IEEEbiography}[{\vspace{6ex}\includegraphics[width=1.1in,height=1.55in,clip,keepaspectratio]{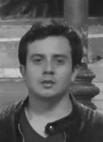}}]{Sebasti\'an Salazar-Colores} received his B. S. degree in Computer Science from Universidad Aut\'onoma Benito Ju\'arez de Oaxaca, received his M. S. degree in Electrical Engineering at Universidad de Guanajuato. He is a PhD candidate in Computer Science at the Universidad Aut\'onoma de Quer\'etaro. His research interest are image processing and computer vision. 
\end{IEEEbiography}

\begin{IEEEbiography}[{\vspace{6ex}\includegraphics[width=1.1in,height=1.55in,clip,keepaspectratio]{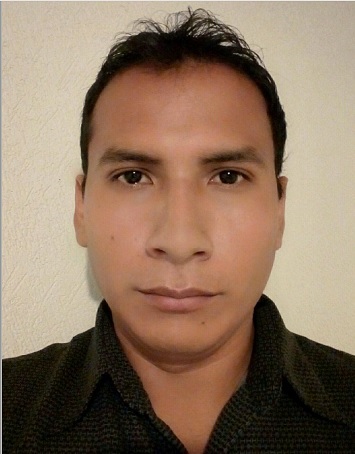}}]{Hugo Moreno Jiménez}received the Robotics Engineering degree in 2016 at the Universidad Polit\'ecnica del Bicentenario in Mex\'ico. Currently is studying his M.S. degree in optomecatronic at Centro de Investigaciones en Optica A.C His research interest is related to digital image and automation.
\end{IEEEbiography}

 \begin{IEEEbiography}[{\vspace{-3ex}\includegraphics[width=1.1in,height=1.45in,clip,keepaspectratio]{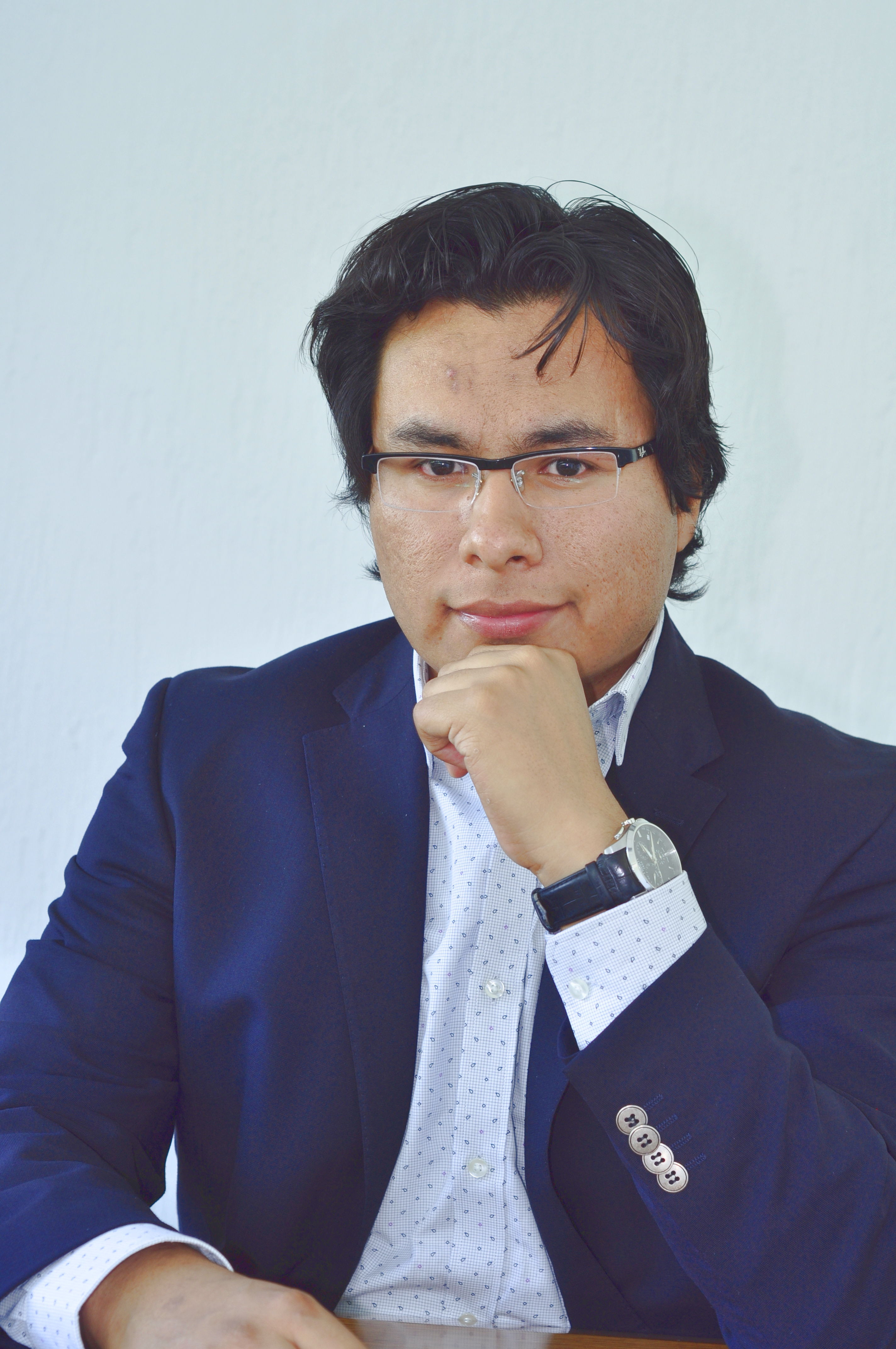}}]
{Gerardo Flores} received the B.S degree in Electronic Engineering with honors from the Instituto Tecnol\'ogico de Saltillo, M\'exico in 2000; the M.S. degree in Automatic Control from CINVESTAV-IPN, Mexico City, in 2010; and the Ph.D. degree in Systems and Information Technology from the Heudiasyc Laboratory of the Universit\'e de Technologie de Compi\`egne - Sorbonne Universit\'es, France in October 2014. From November 2014 to July 2016, he was a post-doctoral researcher with the Centro de Investigaci\'on y de Estudios Avanzados del Instituto Polit\'ecnico Nacional, Mexico City. Since August 2016, he has been full time researcher and Head of the Perception and Robotics LAB with the Center for Research in Optics, Le\'on Guanajuato, Mexico. His research interests are focused on the theoretical and practical problems arising from the development of autonomous robotic systems and vision systems. He is especially interested in: artificial vision, theory and applications of intelligent systems, nonlinear control, design, conception and control of UAVs.Dr. Flores has published 40 papers in the areas of control systems, computer vision and robotics. He has been member of the \textit{Sistema Nacional de Investigadores} since 2014.
\end{IEEEbiography}

\begin{IEEEbiography}[{\vspace{6ex}\includegraphics[width=1.1in,height=1.55in,clip,keepaspectratio]{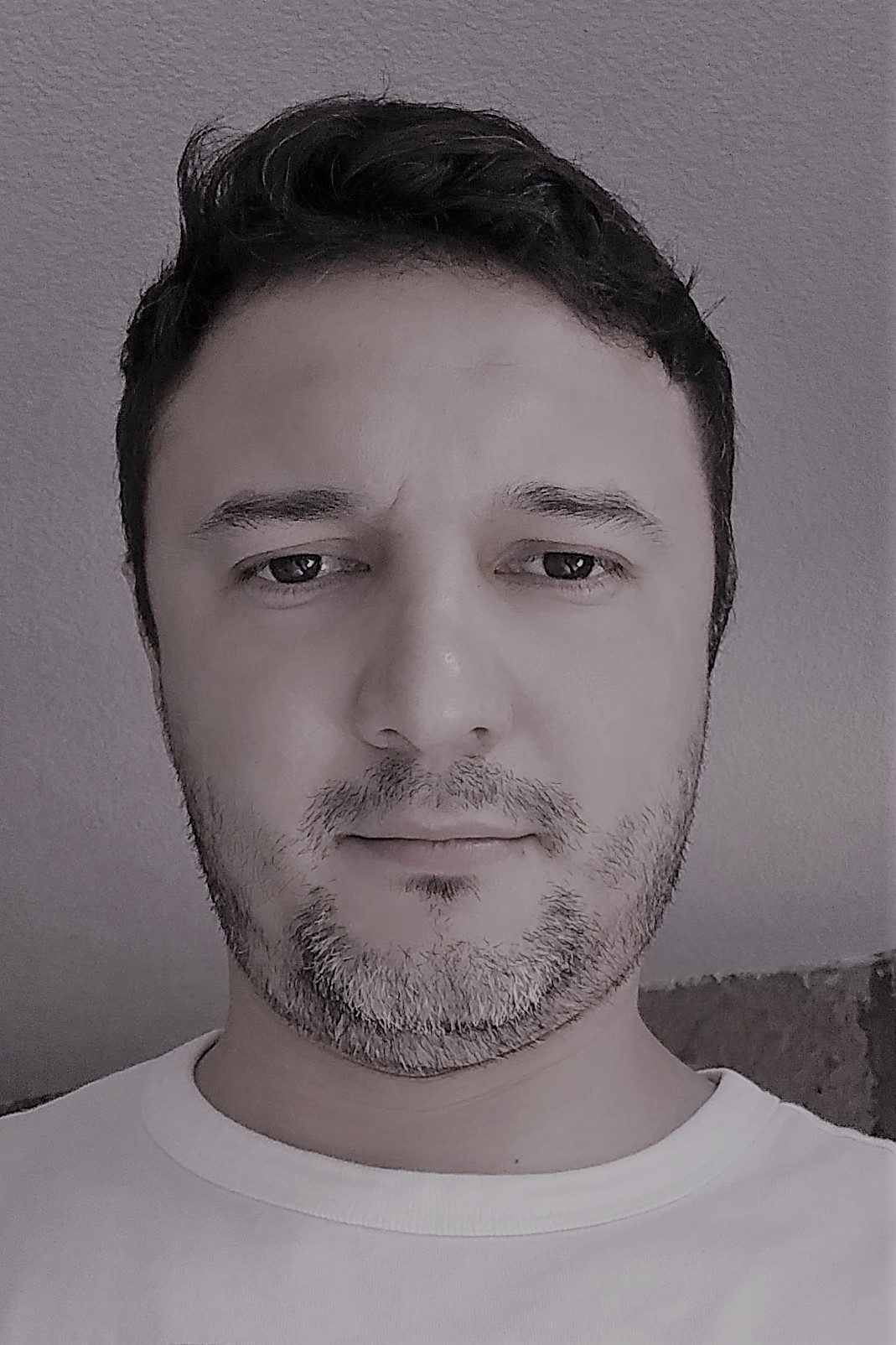}}]{César Javier Ortiz}received the Electronic Engineering grade in 2009 at the Universidad del Quind\'io in Colombia. In 2015 received the M.Sc degree in material science from the same institution. Currently is a Ph.D. candidate of computer science at the Universidad Aut\'onoma de Quer\'etaro. His research interest is related to digital image, and signal processing mainly focused on biomedical applications.
\end{IEEEbiography}

\end{document}